# Temperature dependent dynamics of photoexcited carriers of $Si_2Te_3$ nanowires

Jiyang Chen,[1] Keyue Wu,[2,3] Xiao Shen,[1] Thang Ba Hoang[1,*] and Jingbiao Cui[1]

[1]*Department of Physics and Materials Science, The University of Memphis, Memphis, TN 38152, USA*
[2]*College of Electrical and Photoelectronic Engineering, West Anhui University, Lu'an 237012, Anhui, People's Republic of China*
[3]*Research Center of Atoms Molecules and Optical Applications, West Anhui University, Lu'an 237012, Anhui, People's Republic of China*

[*]Email address: tbhoang@memphis.edu

Abstract: We report an optical study of the dynamics of photoexcited carriers in $Si_2Te_3$ nanowires at various temperatures and excitation powers. $Si_2Te_3$ nanowires were synthesized, by using gold as a catalyst, on a silicon substrate by the chemical vapor deposition method. The photoluminescence spectrum of $Si_2Te_3$ nanowires was primary dominated by defect and surface states related emission at both low and room temperatures. We observed that the decay time of photoexcited carries was very long (> 10 ns) at low temperatures and became shorter (< 2 ns) at room temperature. Further, the carrier decay time became faster at high excitation rates. The acceleration of the photoexcited carrier decay rates indicate the thermal quenching along with the non-radiative recombination at high temperature and excitation power. Our results have quantitatively elucidated decay mechanisms that are important towards understanding and controlling of the electronic states in $Si_2Te_3$ nanostructures for optoelectronic applications.

Recent advances in semiconductor growth techniques have enabled the synthesis of a new class of silicon telluride ($Si_2Te_3$) nanostructures such as nanoplates,[1-3] nanobelts (nanoribbons)[1,4] and nanowires (NWs).[4] $Si_2Te_3$ exhibits interesting structural,[2,5,6] optical,[2,3,7,8] electrical,[9,10] chemical[1,3] and thermal[11,12] properties which promise potential applications ranging from thermo-electrics to optoelectronics. Bulk $Si_2Te_3$ was reported to have an indirect electronic band structure.[7] 2D $Si_2Te_3$ was also being investigated recently and it is interesting that the structural characteristics of this material might vary significantly depending on an array of parameters therefore the resulting band structure is also varying. For instance, the band gap can vary up to 40% along with a 5% change in the lattice constant depending on the orientations of the silicon dimers which locate in between Te atoms.[6] In any case, while the optical, electrical and thermal properties of bulk $Si_2Te_3$ were explored several decades ago, these properties of nanostructured $Si_2Te_3$ have only been examined in the last several years. Indeed, there have been several papers reported on the optical properties of nanoplates[1-3] and nanobelts (nanoribbons)[1,4] but so far none has reported any study on the dynamics of photoexcited carriers in $Si_2Te_3$ nanostructures, which is crucially important information for any application in optoelectronics. One of the challenges in investigating physical properties of nanostructured $Si_2Te_3$ is the stability of the material under ambient conditions. This is because the large surface to volume ratio at the nanoscale that leads to the surface reaction to the water vapor in the atmosphere, resulting a thin Te layer. Furthermore, the complication of the structural properties of $Si_2Te_3$ at low dimensions due to the orientation of the silicon dimers at different temperatures and strain could also lead to strikingly different optical or electronic properties.[6] On the other hands, these dependencies also offer an opportunity such that properties of $Si_2Te_3$ are controllable using these set of parameters. Further, while $Si_2Te_3$ material can be processed using standard semiconductor techniques, it is highly sensitive to environment which offers an advantage for chemical sensing applications.[1,3]

In this work, we study the dynamics of photoexcited carriers in ensembles of $Si_2Te_3$ nanowires as functions of temperature and excitation power. We observed that the photoluminescence (PL) intensity and decay time of photoexcited carries vary significantly under different measurement conditions. Specifically, the emitted photons exhibited a long decay time (> 10.0 ns) at low temperature (< 100 K) but much shorter (~ 1.8 ns) at room temperature, associated with an abrupt reduction in the intensity. Further, at any given temperature, the decay time reduced as a result of the increasing excitation power. These results indicate a significant non-radiative recombination rate associated with defects/surface states. Increased temperature could also change the crystal structure of the nanowires which led to modification of the band structure and the carrier dynamics. Our study offers an insight into the decay dynamics of photoexcited carrier and hints possible applications of the $Si_2Te_3$ nanowires in optoelectronics.

Tellurium (30 mesh, 99.997%, Aldrich) and silicon (325 mesh, 99%, Aldrich) powders were purchased from Sigma Aldrich and used as source materials for $Si_2Te_3$ nanowire growth. Both the Te and the Si powders were placed in a ceramic crucible and loaded into a high temperature tube furnace. The Si or $SiO_2$ substrates were spin coated with a thin gold (Au) film and were placed downstream of gas flow in the furnace. The quartz tube was evacuated by a vacuum pump and was introduced by high purity nitrogen gas (> 99.995%) with a mass flow rate of 15 sccm to maintain the starting pressure at 9.12 Torr. The furnace was heated from the room temperature at 20 ºC/min and maintained at 850 ºC for $Si_2Te_3$ nanowire growth for 5 min. After growth, the ceramic crucible and substrates were then cooled down to room temperature. Further, in order to minimize the surface reaction to the water vapor in the atmosphere, the freshly grown $Si_2Te_3$ nanowires were immediately transferred to a vacuum storage container or an optical cryostat for optical measurements.

In the experimental procedure ensembles of $Si_2Te_3$ nanowires were optically excited by an ultrafast laser (Coherent Chameleon Ultra II, 80 MHz, 150 fs). The output (950 nm) of the laser was frequency doubled (to 475 nm) and focused onto the sample, via an 20X objective lens, with an excitation spot size ~ 3 μm. The average laser excitation power was 100 μW (measured before the objective lens) except for the power dependent experiments when the laser power was varied. $Si_2Te_3$ nanowire sample was mounted onto the cold finger of a closed- cycle cryostat (Janis model CCS-XG-M/204N). The temperature of the cryostat can be controlled to allow optical measurements from 7 – 300 K. The PL signal was collected by the same objective lens and analyzed by a spectrometer (Horiba iHR550) and detected by a CCD (Charged-Coupled Device) camera (Horiba Synapse). For the time-resolved single photon counting measurements, PL signal was filtered by a long pass filter and collected by a fast-timing avalanche photodiode (APD-PDM, Micro Photon Device). A time-correlated single photon counting module (PicoHarp 300) with a time bin of 4 ps was used to analyze the number of photons as a function of time when they arrived at the photodiode. Final lifetimes were obtained from fits to the data de-convolved with the instrument response function.[13]

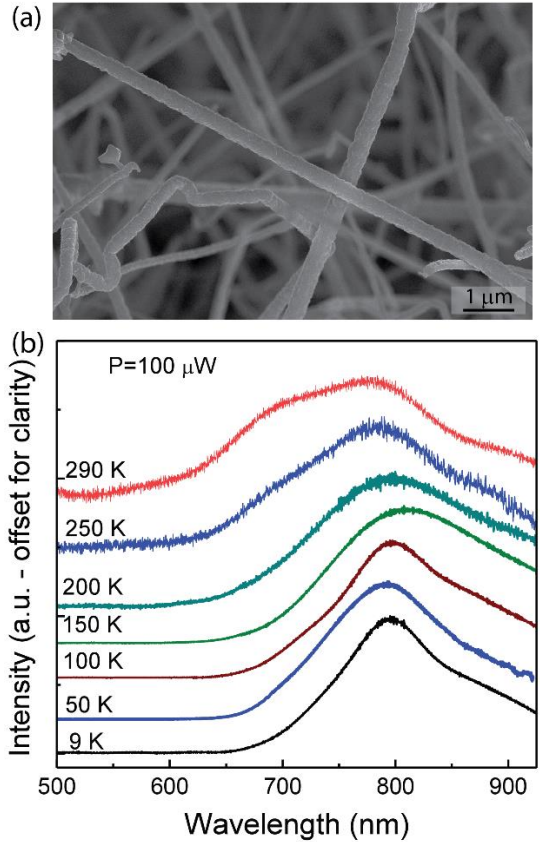

*Figure 1 (a) SEM image of NW ensemble on a silicon substrate. (b) Normalized PL emission spectra of an NW ensemble at various temperatures (9 – 290 K). The spectra were vertical offset for clarity.*

$Si_2Te_3$ NWs were first characterized by the scanning electron microscopic (SEM) method. Figure 1a shows a representative SEM image where as-grown $Si_2Te_3$ NWs were randomly oriented on a silicon substrate. Individual nanowires have diameter and length vary around 300 nm and 10 μm, respectively. Even though most of NWs were straight and relatively uniform in diameter, it appeared that the NWs exhibited rough surfaces which could eventually affect the optical properties as we will discuss below. Further, due to the random orientation of the NW ensembles, in the optical experiments we have circularly polarized the laser excitation so many NWs were somewhat equally excited in term of polarization. Figure 1b shows the normalized PL emission spectra of an NW ensemble at various temperatures (9 – 290 K) at a fixed laser excitation power $P$ =100 μW. At low temperature (< 100 K) the PL spectrum featured by a broad peak at around 790 nm with a full-half-width-maximum (FHWM) ~ 90 nm and a shoulder at longer wavelength (~ 890 nm). Similar results were observed earlier by Wu et al.[2] for $Si_2Te_3$ nanoplates where these two emission bands were assigned as defect emissions associated with thermal quenching as the temperature increased. It is however noted that in this work we did not observe the band gap PL emission near 563 nm (2.2 eV for bulk $Si_2Te_3$).[2,7,8,11] This could be due to a different excitation conditions (pulsed vs continuous laser),[14] crystal quality or could be due to a modification of the band structure in these NWs caused by the reorientation of the Si dimers.[6] We also note that some PL emission bands from bulk $Si_2Te_3$ were previously observed to exhibit at very low energies (~ 1.1 eV or > 1100 nm) due to the recombination at trap states above the valence band.[11] These emission bands were not detected in our experiment due to a limitation of the silicon CCD and other optics. At higher temperatures, another shoulder appeared at around 700 nm and became more dominant at 290 K. This additional shoulder was likely originated from defect centers that located at higher energy levels (within the band gap) and only be populated at high temperatures or high excitation powers. The PL emission at various temperatures is further analyzed in Fig. 2 which displays the temperature dependence of integrated PL intensity at temperatures from 9 K to 290 K. At low temperature (< 75 K) the total emitted intensity is high and fairly a constant but at above 75 K the intensity abruptly reduced. This is similar to previously observed behavior for $Si_2Te_3$ nanoplates.[2] The inset in Fig. 2 displays the laser excitation power dependence of the total integrated PL intensity at temperature T = 9 K. It is clear that below certain excitation value (< 250 μW) the PL intensity increased linearly with the increasing excitation power. Above 250 μW, a saturation behavior occurred which was a result of the reduction in the absorption efficiency. It was also possible that at high excitation powers, the high average number of photoexcited carriers led to the non-radiative recombination with the surface states and resulted in nonlinear relaxation processes (i.e. Auger recombination) which are characterized by a reduction in the photon decay time.

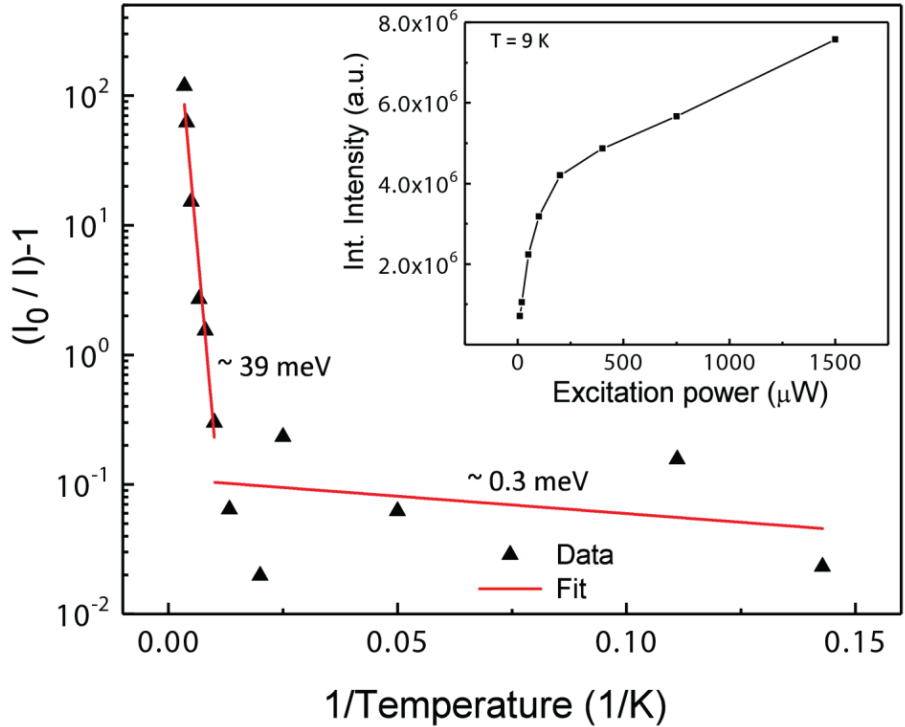

*Figure 2 Temperature dependent integrated PL intensity at a fixed average excitation power P =100 µW. I and I₀ are the PL intensities at a given temperatures T and at 0K, respectively. Red lines are fits to the data. Inset: Excitation power dependent PL emission intensity at temperature T = 9 K.*

In the previous study by Wu et al,[2] by measuring the PL spectra at different temperatures the authors have extracted very large activation energies which indicated the several thermal related processes have occurred. Here we observed also a large activation energy of $E = 39$ meV at temperature above 75 K. The red lines in Figures 2 were the fits according to $I_0/I - 1 = exp(-E/k_BT)$ where I and I₀ are the PL intensities at a given temperatures T and at 0K, respectively. This activation energy is even larger than the value of the thermal energy $k_BT$ at room temperature (~ 25.7 meV) which suggests that the energy transition of the photoexcited carriers was not purely promoted by the thermal energy but rather a thermal-induced modification of the defect related state energy levels. For instance, defects related to impurities, rough surfaces (Fig. 1(a)) or structural/chemical modifications could transform to metastable configurations with different energy levels under thermal and optical excitations. A well-known case is the DX center in III-V semiconductors.[15] Such metastable configurations can affect the decay of photoexcited carriers differently and their populations depend on temperature and excitation rate. This hypothesis is further supported by the decay dynamic measurements of the photoexcited carrier as we will present in sections below.

In order to gain further insight into the dynamics of the photoexcited carriers we performed the time-resolved measurements (Figure 3). At low temperatures (< 100 K) and low excitation power (< 100 µW), the decay time of the photoexcited carriers was characterized by a straight line, indicating a long decay time of the excitonic states. Indeed, given 80 MHz of the laser used for this study, we could not perform a fitting procedure to deduct the long decay times for these measurement conditions. This result is in consistent with a previous study which concluded a long hole's lifetime in bulk $Si_2Te_3$.[11] At around 150 K, a decay trace could be observed with a single decay component that has a decay time ~ 10 ns at 100 µW excitation power. At higher temperatures, the decay time became faster and at 290 K it was determined to be 1.83 ns. A significant reduction of the decay time when the temperature increased has indicated several effects. First, the thermalization process of neutral donors and thermal quenching of the photoexcited. This is in consistent with the observation that there was a significant reduction in the total integrated PL emission intensity (Fig. 2). The thermal processes led to an increase in the non-radiative recombination rate and eventually a reduction in the decay time of the emitted photons. Secondly, following the work by Shen *et al*. there could be a re-arrangement in the structural configuration of the $Si_2Te_3$ lattice at different temperatures which resulted in a modification in the band structure and carrier dynamics.[6]

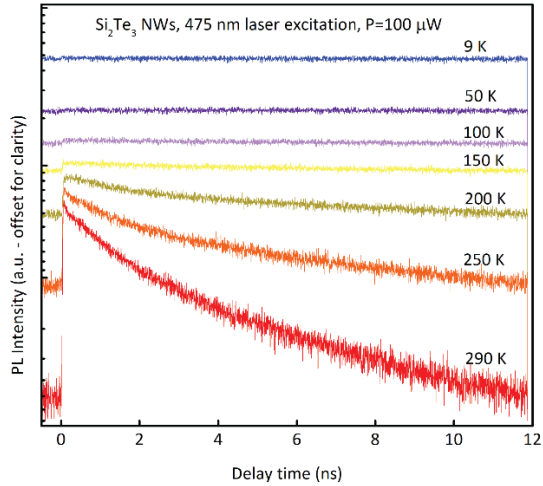

*Figure 3 Measured decay curves at different temperatures at a fixed average excitation power P = 100 μW.*

The decay dynamics of the photoexcited carriers in $Si_2Te_3$ nanowires could also be associated with the carrier generation rate, which is directly related to the incident laser excitation power. Figure 4 shows the decay dynamics of $Si_2Te_3$ NW ensembles in a matrix of parameters. Specifically, we have measured the decay times of these NWs at various incident laser excitation powers and temperatures. At low excitation powers (< 100 μW) and low temperatures (< 100 K), the decay curves were featured by flat lines from which the decay time could not be determined at least from our current apparatus. At below 100 K, by increasing the excitation power (to above 150 μW), a single exponential decay component could be observed. For instance, at 9 K the decay time was determined to decrease from 7.25 ns to 6.08 ns when the excitation power increase from 200 to 750 μW (Fig. 4(a)). Other similar measurements at temperatures 50-290 K are shown in Figs. 4(b)-(f). At room temperature (290 K), the decay time reduced from 1.88 ns to 1.73 ns as the excitation power increased from 20 – 750 µW. The temperature and power excitation dependences imply that, even at low temperature, the high excitation rate has led to a non-radiative recombination of the photoexcited carries (due to surface states, for instance). At higher temperature (> 150 K), the decay time became faster even at low excitation rates which was the result of thermal effect.

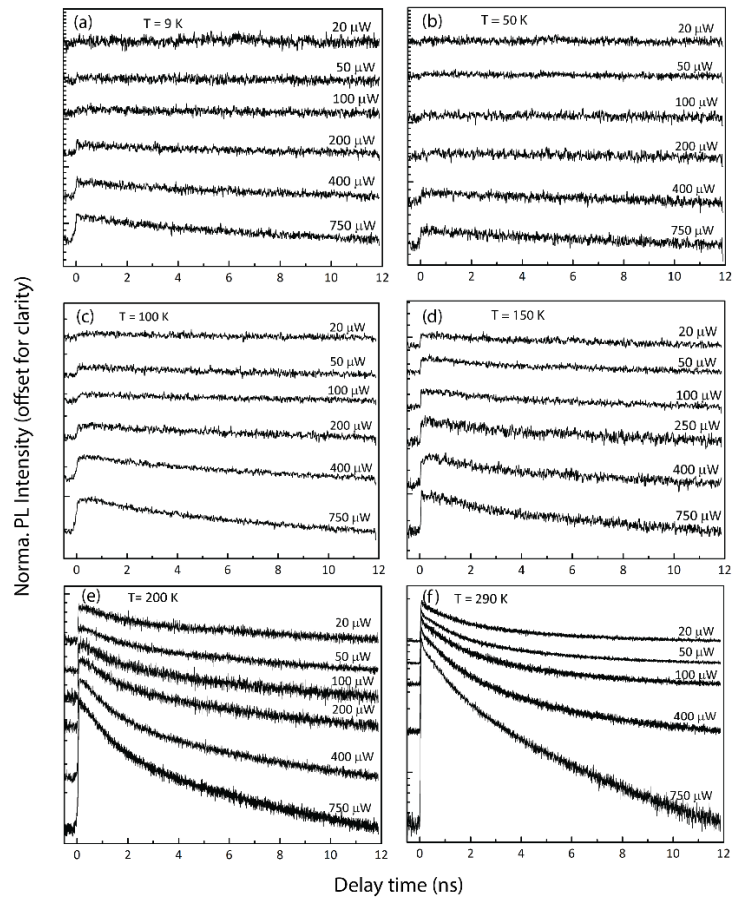

*Figure 4 Excitation power dependent decay dynamics at different temperatures (a-f) → (9-290 K). At each temperature, the excitation power was varied from 20 µW to 750 µW.*

One another possible mechanism which could lead to the shorter decay time at higher temperature and excitation rate is due to the modification of the band structure as the result of changes in the orientation of the Si dimers with respect to the Te atoms.[6] This could be further complicated by the fact that there were boundaries between different crystal domains, which may have different orientation of the dimers. The energy offsets between domains, depending on the temperature, could result in an indirect transition and therefore a long decay time. The current available data achieved in this study did not allow us to draw a direct conclusion regarding a specific band alignment but rather hinted several mechanisms for the decay dynamics of the photoexcited carries. Future measurements such as polarized decay dynamics from single $Si_2Te_3$ NWs in different excitation conditions, and in combination with structural characterizations (SEM or TEM – Transmission Electron Microscopy) to correlate morphological characteristics with optical properties of the same single NWs could potentially provide further information.

In conclusion, we have investigated the decay dynamics of the photoexcited carries in ensembles of $Si_2Te_3$ NWs as functions of temperature and excitation rate. We have observed a combination of carrier thermalization and possible band structure modification. Our results have revealed the decay dynamics of $Si_2Te_3$ NWs in a matrix of parameters which could be used to control the physical characteristic of the materials for possible applications in optoelectronics and thermoelectric.

Acknowledgement

This work was supported by the National Science Foundation (NSF) (DMR-1709528 and DMR-1709612), by the Ralph E. Powe Junior Faculty Enhancement Award from Oak Ridge Associated Universities (to XS), and in part by a grant from The University of Memphis College of Arts and Sciences Research Grant Fund (to TBH). This support does not necessarily imply endorsement by the University of research conclusions.

References

1       Sean Keuleyan, Mengjing Wang, Frank R. Chung, Jeffrey Commons, and Kristie J. Koski,  Nano Letters **15** (4), 2285 (2015).
2       Keyue Wu, Weiwei Sun, Yan Jiang, Jiyang Chen, Li Li, Chunbin Cao, Shiwei Shi, Xiao Shen, and Jingbiao Cui,  Journal of Applied Physics **122** (7), 075701 (2017).
3       Mengjing Wang, Gabriella Lahti, David Williams, and Kristie J. Koski,  ACS Nano **12** (6), 6163 (2018).
4       Keyue Wu and Jingbiao Cui,  Journal of Materials Science: Materials in Electronics (2018).
5       K. Ploog, W. Stetter, A. Nowitzki, and E. Schönherr,  Materials Research Bulletin **11** (9), 1147 (1976).
6       X. Shen, Y. S. Puzyrev, C. Combs, and S. T. Pantelides,  Applied Physics Letters **109** (11), 113104 (2016).
7       A. P. Lambros and N. A. Economou,  physica status solidi (b) **57** (2), 793 (1973).
8       U. Zwick and K. H. Rieder,  Zeitschrift für Physik B Condensed Matter **25** (4), 319 (1976).
9       H. P. Bauer and U. Birkholz,  physica status solidi (a) **49** (1), 127 (1978).
10      M. Rick, J. Rosenzweig, and U. Birkholz,  physica status solidi (a) **83** (2), K183 (1984).
11      K. Ziegler and U. Birkholz,  physica status solidi (a) **39** (2), 467 (1977).
12      Rinkle Juneja, Tribhuwan Pandey, and Abhishek K. Singh,  Chemistry of Materials **29** (8), 3723 (2017).
13      Jörg Enderlein and Rainer Erdmann,  Optics Communications **134** (1), 371 (1997).
14      Jiani Huang, Thang B. Hoang, and Maiken H. Mikkelsen,  Scientific Reports **6**, 22414 (2016).
15      P. M. Mooney, in *Deep Centers in Semiconductors: A State-of-the-Art Approach*, edited by Sokrates T. Pantelides (CRC Press, 1992), pp. 643.